\documentclass[prl,twocolumn,showpacs,superscriptaddress]{revtex4-1}

\usepackage[applemac]{inputenc}
\usepackage[T1]{fontenc}
\usepackage{color}
\usepackage{amsmath}
\usepackage{amssymb}
\usepackage{graphicx}
\usepackage{float}

\begin{document}


\newcommand{\bra}[1]{\langle{#1}|}
\newcommand{\ket}[1]{|{#1}\rangle}
\newcommand{\id}{\openone}
\newcommand{\eq}[1]{(\ref{#1})}
\newcommand{\vphi}{\varphi}
\renewcommand{\vec}[1]{\mathbf{#1}}
\newcommand{\bo}{\boldsymbol{(}}
\newcommand{\bc}{\boldsymbol{)}}

\newcommand{\sz}[1]{\sigma_{z}^{#1}}
\newcommand{\sx}[1]{\sigma_{x}^{#1}}
\newcommand{\sy}[1]{\sigma_{y}^{#1}}
\newcommand{\spp}[1]{\sigma_{+}^{#1}}
\newcommand{\smm}[1]{\sigma_{- }^{#1}}

\newcommand{\tsz}[1]{\tilde\sigma_{z_{#1}}}
\newcommand{\tsx}[1]{\tilde\sigma_{x_{#1}}}
\newcommand{\tsy}[1]{\tilde\sigma_{y_{#1}}}
\newcommand{\tspp}[1]{\tilde\sigma_{+_{#1}}}
\newcommand{\tsmm}[1]{\tilde\sigma_{-_{#1}}}

\newcommand{\sch}{Schr\"odinger}
\newcommand{\schs}{Schr\"odinger's}
\newcommand{\nn}{\nonumber}
\newcommand{\nl}{\nn \\ &&}
\newcommand{\dg}{^\dagger}
\newcommand{\rt}[1]{\sqrt{#1}\,}
\newcommand{\ito}{It\^o }
\newcommand{\str}{Stratonovich }
\newcommand{\erf}[1]{Eq.~(\ref{#1})}
\newcommand{\erfs}[2]{Eqs.~(\ref{#1}) and (\ref{#2})}
\newcommand{\erft}[2]{Eqs.~(\ref{#1}) -- (\ref{#2})}

\newcommand{\im}{\mathrm{Im}}

\newcommand{\wa}[1]{\omega_{a_{#1}}}
\newcommand{\twa}[1]{\tilde{\omega}_{a_{#1}}}
\newcommand{\wrrb}{\omega_{r_{b}}}
\newcommand{\wrr}{\omega_{r}}
\newcommand{\rf}{\mathrm{rf}}
\newcommand{\spec}{\mathrm{s}}

\newcommand{\w}[1]{\omega_{#1}}
\newcommand{\wdr}[1]{\omega_{d_{#1}}}

\newcommand{\ad}{a^\dag}
\newcommand{\aop}{a}

\newcommand{\veps}{\varepsilon}

\newcommand{\ccos}{\mathrm{c}}
\newcommand{\ssin}{\mathrm{s}}
\newcommand{\ej}[1]{E_\mathrm{J_{#1}}}
\newcommand{\ec}[1]{E_\mathrm{C_{#1}}}
\newcommand{\eel}[1]{E_\mathrm{el_{#1}}}
\newcommand{\vlc}{V_\mathrm{LC}}

\newcommand{\nc}{n_\mathrm{crit}}

\newcommand{\etal}{{\it et. al}}

\newcommand{\U}[1]{{\mathsf{#1}}}
\newcommand{\rhos}{{\rho_S}}
\newcommand{\drhos}{{\dot\rho_S}}

\newcommand{\red}{\color[rgb]{0.8,0,0}}
\newcommand{\green}{\color[rgb]{0.0,0.6,0.0}}
\newcommand{\blue}{\color[rgb]{0.0,0.0,0.6}}
\newcommand{\jcomment}[1]{\emph{\color[rgb]{0,0,0.6}(#1)}}
\newcommand{\kcomment}[1]{\emph{\color[rgb]{0,0.6,0}(#1)}}

\def\be{\begin{equation}}
\def\ee{\end{equation}}

\title{Tunable joint measurements in the dispersive regime of cavity QED}
\date{\today}

\author{Kevin Lalumi\`ere}
\affiliation{D\'epartement de Physique, Universit\'e de Sherbrooke, Sherbrooke, Qu\'ebec, Canada, J1K 2R1}
\author{J. M. Gambetta}
\affiliation{Institute for Quantum Computing and Department of Physics and Astronomy, University of Waterloo, Waterloo, Ontario, Canada, N2L 3G1}
\author{Alexandre Blais}
\affiliation{D\'epartement de Physique, Universit\'e de Sherbrooke, Sherbrooke, Qu\'ebec, Canada, J1K 2R1}

\begin{abstract}
Joint measurements of multiple qubits open new possibilities for quantum information processing.  Here, we present an approach based on homodyne detection to realize such measurements in the dispersive regime of cavity/circuit QED.  By changing details of the measurement, the readout can be tuned from extracting only single-qubit to only multi-qubit properties. We obtain a reduced stochastic master equation describing this measurement and its effect on the qubits. As an example, we present results showing parity measurements of two qubits. In this situation, measurement of an initially unentangled state can yield with near unit probability a state of significant concurrence.
\end{abstract}


\maketitle

In most of the current quantum information experiments, measurements are used to extract information only about single-qubit properties.   Joint measurements where information about  both single and multi-qubit properties can be  obtained offer new possibilities. Examples are the test of quantum paradoxes~\cite{lundeen:2009a}, test of quantum contextuality~\cite{kirchmair:2009a}, realization of quantum state tomography with weak measurements~\cite{filipp:2008a,DiCarlo:2009,chow:2009a} and cluster state preparation~\cite{louis:2007a}.  A particularly powerful type of joint measurement is parity measurement, where information is gained only about the overall parity of the multi-qubit state, without any single-qubit information.  This type of measurement can be used for the generation of entanglement without unitary dynamics~\cite{mao:2004a,trauzettel:2006a,williams:2008a,hill:2008a}, for quantum error correction~\cite{bacon:2006a,kerckhoff:2009a}, and deterministic quantum computation with fermions~\cite{beenakker:2004a,engel:2005a}. In this paper, we show how such joint measurements can be realized in the dispersive regime of cavity QED~\cite{haroche:2006a}.  In particular, we show how the character of the measurement can be tuned from purely single-qubit to parity readout.  As a realistic example, we present results for circuit QED~\cite{filipp:2008a,DiCarlo:2009,chow:2009a} and show that states with large concurrence can be obtained.  Entanglement generation by measurement was previously studied in this system~\cite{Hutchison:2008,Rodrigues:2008,Helmer:2009,Bishop:2009}, but ignoring information about the parity.  With parity measurements, entanglement generation by measurement can be deterministic rather than probabilistic.

We consider a pair of two-level systems (i.e. qubits) of frequencies $\wa{j}$ with $j=1,2$ coupled to a high-Q cavity of frequency $\omega_r$.  In the dispersive limit, where $|\Delta_j|=|(\wa{j}-\wrr)|\gg|g_j|$ with $g_j$ the coupling strength of qubit $j$ to the cavity, the Hamiltonian of this system takes the form~\cite{Blais:2007}
\begin{equation}\label{eq:H}
\begin{split}
H=
&(\wrr+\sum_j \chi^{j} \sz{j}) \ad \aop + \sum_{j} \frac{\tilde{\omega}_{a_j}}{2}\sz{j}
+ J_{q}(\smm{1}\spp{2}+\smm{2}\spp{1})
\\&+\epsilon_m(t) (\ad e^{-i\omega_m t}+\mathrm{h.c.}).
\end{split}
\end{equation}
This result is valid to second order in the small parameter $\lambda_j = g_j/\Delta_j$. Here, we have defined the dispersive coupling strength $\chi^{j} = g_j\lambda_j$, the Lamb-shifted qubit frequency $\tilde{\omega}_{a_j}$ and the strength of qubit-qubit coupling mediated by virtual photons $J_q=g_1 g_2 (1/\Delta_1+1/\Delta_2)/2$~\cite{Blais:2007}.  The last term represents a coherent drive on the cavity of amplitude $\epsilon_m(t)$ and frequency $\omega_m \approx \omega_r$, appropriate for measurement of the qubits.  With this choice of drive frequency, we have safely dropped a qubit driving term of amplitude $\lambda_{j} \epsilon_m$~\cite{Blais:2007}.  In order to focus on entanglement generated by measurement only, we drop the term proportional to $J_q$.  This is reasonable since the possible measurement outcomes are eigenstate of the flip-flop interaction $\smm{1}\spp{2}+\smm{2}\spp{1}$, as will be clear below.

Coupling to unwanted degrees of freedom is modeled by using a Lindblad-type master equation~\cite{walls:2008a}.
In Ref.~\cite{Hutchison:2008}, a master equation for the qubits only was obtained by enslaving the cavity to
the qubit dynamics. This approach is valid only in the limit where damping of the cavity $\kappa$
greatly overwhelms the dispersive coupling strength $\chi^j$.  Here, we go beyond these results by using a polaron-type
transformation to trace-out the cavity~\cite{Gambetta:2008,boissonneault:2009a}.  Starting from Eq.~\eq{eq:H}, we find  following Ref.~\cite{Gambetta:2008} the effective master equation
\begin{equation}\label{eq:dotrho}
\begin{split}
\dot{\rho}
\approx&
- i[\sum_j \frac{\tilde{\omega}_{a_j}}{2} \sz j,\rho]
+\sum_j \gamma_{1j} \mathcal{D}[\smm{j}]\rho
+\sum_j \frac{\gamma_{\phi j}}{2} \mathcal{D}[\sz{j}]\rho
\\&
+\kappa \mathcal D[\sum_j \lambda_j \smm j]\rho
+\sum_{xy} \left( \Gamma_d^{xy}-i A_c^{xy} \right) \Pi_x \rho \Pi_y
\equiv\mathcal L\rho,
\end{split}
\end{equation}
where $\mathcal D[c] \cdot = c\cdot c\dg-\{c\dg c,\cdot\}/2$. In this expression, $\gamma_{1j}$ is the relaxation rate of qubit $j$ and $\gamma_{\phi_j}$ its pure dephasing rate. The fourth term represents Purcell damping at the rate $\lambda_j^2\kappa$~\cite{boissonneault:2009a}, while the last contains both measurement-induced dephasing ($\Gamma_d^{xy}$) and ac-Stark shift ($A_c^{xy}$) by the measurement photons~\footnote{In the single qubit case, this last term reduces to Eqs.~(3.11) and (3.13) of Ref.~\cite{Gambetta:2008}.}. In Eq.~\eq{eq:dotrho}, $x$ ($y$) stands for one of the four logical states $ij$ with $i,j\in\{g,e\}$ the qubit's ground and excited states and $\Pi_x=\ket{x}\bra{x}$.  Measurement-induced dephasing and ac-Stark shift are given by
\be
\Gamma_d^{xy}=(\chi_x-\chi_y)\mathrm{Im}[\alpha_x \alpha_y^*],
\ee
\be
A_c^{xy}=(\chi_x-\chi_y)\mathrm{Re}[\alpha_x \alpha_y^*],
\ee
where $\chi_x=\bra{x}\sum_j \chi^{j} \sz{j}\ket{x}$ and $\alpha_x$ the amplitude of the coherent state when the qubits are in state $|x\rangle$.  This amplitude satisfies
\be
\dot{\alpha}_x=-i(\wrr +\chi_x) \alpha_x -i \epsilon_m(t) e^{-i\omega_m t} - \kappa \alpha_x/2.
\ee
The reduced master equation Eq.~\eq{eq:dotrho} is a very good approximation to the full dynamics when $\kappa/2 \gg  \gamma_{1j}$. Since $\gamma_{1j}$ does not include Purcell damping, this inequality is easily satisfied with current Purcell limited qubits~\cite{houck:2008a}.

\begin{figure}[t]
\centering \includegraphics[width=\columnwidth]{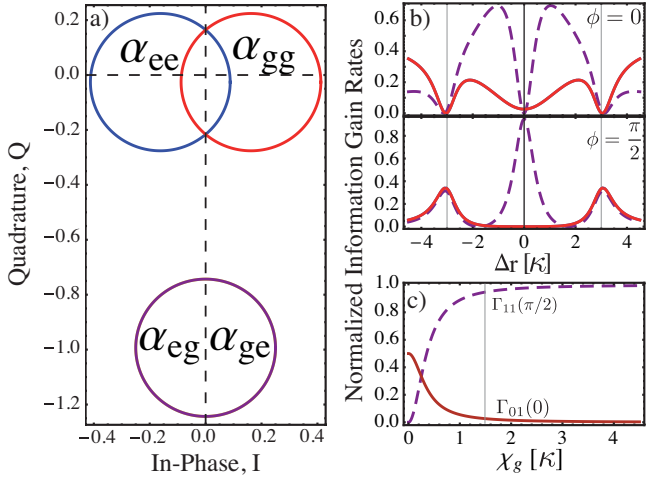}
\caption{(Color online)
a) Phase space illustration of the stationary states $\ket{\alpha_{ij}}$ for: $g_1=-g_2=-  15\kappa$ and $\chi^{j} \sim 1.5 \kappa$.  The drive is at resonance with the bare cavity $\Delta_r=0$ and its amplitude is $\epsilon=\kappa/2$.
b) Normalized rates of information gain for $\phi=0$ or $\phi=\pi/2$:  $\Gamma_{01}(\phi)=\Gamma_{10}(\phi)$ (full red line), $\Gamma_{11}(\phi)$ (dashed purple line). The vertical lines are at $\pm2\chi^j$.
c) Normalized rates at $\Delta_r=0$ as a function of $\chi^j$: $\Gamma_{01}(0)=\Gamma_{10}(0)$ (full red line), $\Gamma_{11}(\pi/2)$ (dashed purple line). The vertical line indicates the value of $\chi^j$ used in panel b). Other rates $\Gamma_{11}(0)$, $\Gamma_{01}(\pi/2)$, and $\Gamma_{10}(\pi/2)$ are zero and not shown.
}
\label{fig:coherent_rates}
\end{figure}

To go beyond information about average evolution, we use quantum trajectory theory of homodyne measurement on the transmitted cavity field to obtain information about single experimental runs~\cite{wiseman:1993a}. Following the approach of Ref.~\cite{Gambetta:2008}, we find in the multi-qubit case the reduced stochastic master equation (SME)
\begin{equation}\label{eq:SME}
\begin{split}
\dot \rho_J
=
&
\mathcal L\rho_J
+ \mathcal{M}[c_\phi]\rho_J \xi(t)-i [c_{\phi-\pi/2},\rho_J]\xi(t)/2,
\end{split}
\end{equation} and the measured homodyne current is proportional to  $J(t)=\mathrm{Tr}[c_\phi\rho_J] + \xi(t)$. Here $\mathcal M[c] \cdot=\{c,\cdot\}/2-\mathrm{Tr}[c\cdot]\cdot$, and $\xi(t)$ is Gaussian white noise satisfying
$E[\xi(t)] = 0$ and $E[\xi(t)\xi(t')] = \delta(t-t')$, with $E[\cdot]$ denoting an ensemble average over realizations of the noise. This stochastic equation is valid for $\kappa/2\gg \gamma_{11}+\gamma_{12}$, which is again easily satisfied~\cite{houck:2008a}.

In Eq.~\eq{eq:SME}, the joint measurement operator $c_\phi$ is
\be
\begin{split}
  c_\phi & =  \sqrt{\Gamma_{10}(\phi)} \sz1+ \sqrt{\Gamma_{01}(\phi)} \sz2 + \sqrt{\Gamma_{11}(\phi)} \sz1\sz2,
\end{split}
\ee
where
\be
\begin{split}
&\Gamma_{ij}(\phi)= \kappa \eta |\beta_{ij}|^2\cos^2(\phi-\theta_{\beta_{ij}}),
\\
&\beta_{ij} =\left(\alpha_{ee}+(-1)^{j}\alpha_{eg}+(-1)^{i}\alpha_{ge}+(-1)^{i+j}\alpha_{gg}\right)/2,
\end{split}
\ee
with $\phi$ the phase of the local oscillator, $\theta_\alpha=\textrm{Arg}(\alpha)$, and $\eta$ the efficiency with which the photons leaking out of the cavity are detected.  $\Gamma_{ij}$ represents the rate of information gained about the first qubit polarization ($ij$=10), second qubit polarization ($ij$=01) or the parity ($ij$=11). An optimal measurement occurs when $c_{\phi-\pi/2}=0$ since, in this case, all the back-action arising from the measurement is associated with information gain~\cite{Gambetta:2008}. Given the form of $c_{\phi-\pi/2}$, this cannot be realized, except in trivial cases.

Given that $\chi^j$, $\Delta_r=\omega_r-\omega_m$ and $\phi$ can be changed in-situ~\cite{filipp:2008a,DiCarlo:2009,chow:2009a}, the form of the measurement operator $c_\phi$ can be tuned (in the dispersive approximation, changing $\epsilon_m$ only leads to an overall rescaling).  There are several useful choices of $c_\phi$.  For example, an equally weighted joint measurement (all $|\Gamma_{ij}|$ equal) is ideal for quantum state tomography since in this case both the required single and two-qubit information are on an equal footing. In the limit $|\chi^1\pm\chi^2|\gg\kappa$, this is achieved by choosing $\omega_m$ to match one of the four pulled cavity frequencies $\omega_r+\chi_x$. As can be seen in Fig.~\ref{fig:coherent_rates}b), for $\chi^1=\chi^2$, an equally weighted joint measurement is realized by setting $\Delta_r=\pm2\chi^j$.  For this choice of $\chi^j$ however, at $\Delta_r=0$ it is not possible to determine which qubit is excited and as a result the measurement is either completely collective ($\sigma_z^1+\sigma_z^2$) for $\phi=0$~\cite{Hutchison:2008} or more interestingly extracts information only about the parity $(\sigma_z^1\sigma_z^2)$ of the combined two-qubit state for $\phi=\pi/2$.

This can be understood by considering the steady-state cavity amplitude $\alpha_x$.  Fig.~\ref{fig:coherent_rates}a) shows a phase-space plot corresponding to the four coherent states $\ket{\alpha_x}$ for the parameters given in the caption. Since $\chi^1=\chi^2$, the coherent states $\alpha_{eg}$ and $\alpha_{ge}$ overlap, while $\mathrm{Im}[\alpha_{ee}]=\mathrm{Im}[\alpha_{gg}]$ but $\mathrm{Re}[\alpha_{ee}]\neq\mathrm{Re}[\alpha_{gg}]$~\footnote{It is possible to interchange the role of $\{ee,gg\}$ and $\{eg,ge\}$ by working with $\chi^1=-\chi^2$}. As a result, measurement of the $Q$ $(\phi=\pi/2)$ quadrature reveals information only about the {\em parity} and $I$ $(\phi=0)$ the collective polarization. Since, for these parameters, there is information in the quadrature orthogonal to the measurement, $c_{\phi-\pi/2}\neq0$, and this measurement is not optimal. As illustrated in Fig.~\ref{fig:coherent_rates}c) however, as the ratio $\chi^j/\kappa$ is increased, the measurement becomes optimal for parity with $\Gamma_{01}(0)$, normalized by $\sum_{ij=01,10,11} [\Gamma_{ij}(0) + \Gamma_{ij}(\pi/2)]$, scaling as  $(\kappa/\chi^j)^2$.

\begin{figure}[t]
\centering \includegraphics[width=\columnwidth]{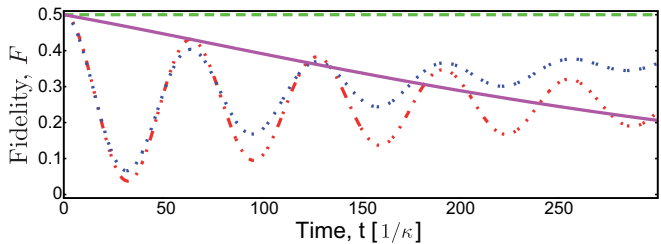}
\caption{(Color online)
Mean fidelity to  $\ket{\phi_+}$ (dashed green and full purple lines) and $\ket{\psi_+}$ (dotted-dashed red and dotted blue lines) obtained from solving Eq.~\eq{eq:dotrho}.
Dashed green and dotted-dashed red lines: $(\gamma_{1j},\gamma_{\phi j}) = (0,0)$.  Full purple and dotted blue lines: ($\kappa/250$,0).
The measurement drive is $\epsilon \tanh(t/\sigma)$ with $\sigma=1/\kappa$ and the initial state $(\ket{g}+\ket{e})\otimes(\ket{g}+\ket{e})/2$. The other parameters are $g_1=-g_2=-  100\kappa$, $\chi^{j} = 10 \kappa$, $\epsilon=\kappa$ and $\kappa/2\pi=5$ MHz.
}
\label{fig:fidelities}
\end{figure}

An application of parity measurements is the generation of entangled states from separable ones~\cite{mao:2004a,trauzettel:2006a,williams:2008a,hill:2008a}.  In contrast to collective polarization measurements~\cite{Hutchison:2008,Rodrigues:2008,Helmer:2009,Bishop:2009}, this can be achieved with unit probability. For example, with the initial separable state $(\ket{g}+\ket{e})\otimes(\ket{g}+\ket{e})/2$, the measurement ideally projects on the Bell states $\ket{\phi_+}=(\ket{eg}+\ket{ge})/\sqrt 2$ or $\ket{\psi_+}=(\ket{gg}+\ket{ee})/\sqrt 2$. That is, evolution under Eq. \eqref{eq:SME} shows a collapse of the separable state to $\ket{\phi_+}$ or $\ket{\psi_+}$, conditioned on the record $J(t)$ being predominately negative or positive respectively.

There are four main causes of errors in this collapse.  The first is relaxation and damping [the dissipative terms of Eq. \eqref{eq:dotrho}].  Interestingly, with the parameters of Fig.~\ref{fig:coherent_rates}, $\lambda_1=-\lambda_2$ such that $\ket{\phi_+}$ is immune from Purcell decay~\cite{Hutchison:2008}. The second is the time-dependent ac-Stark shift [unitary contribution from the last term of Eq.~\eqref{eq:dotrho}] which causes a phase accumulation between $\ket{gg}$ and $\ket{ee}$ in $\ket{\psi_+}$.  This contribution can be seen as a slow oscillation of the fidelity $F=\bra{\psi} \rho \ket{\psi}$ between the state $\ket{\psi_+}$ and those obtained by numerical integration of Eq.~\eq{eq:dotrho}. This is illustrated in  Fig.~\ref{fig:fidelities}.  There, the mean fidelity to $\ket{\phi_+}$ is always $1/2$, since half the density matrices collapse to that state, while oscillations due to the ac-Stark shift appears in the fidelity to $\ket{\psi_+}$. However, this shift is deterministic and can thus be undone.  The third error comes from $c_{0}\neq 0$ causing a stochastic phase between $\ket{gg}$ and $\ket{ee}$ [last term of Eq.~\eqref{eq:SME}].  For a given experimental run, this does not reduce the concurrence or purity of the state [because $\xi(t)$ is known from  $J(t)$]. However, since this phase varies from shot to shot, the ensemble averaged state is mixed. This error can be overcomed by performing $J(t)$-dependent single qubit phase operations after the measurement or, more simply, by operating in the large $\chi^j$ limit where its effect is negligible as illustrated in Fig.~\ref{fig:coherent_rates}c).  Finally, the measurement is not ideal in the sense that measurement-induced dephasing affects the measurement outcome $\ket
{\psi_+}$ (i.e.~$\Gamma_d^{ee,gg} \neq 0$).  However, this effect can be made negligible by increasing the ratio  $\chi^j/\kappa$ since $\Gamma_{11}(\pi/2)/\Gamma_d^{ee,gg} \sim ( \chi^j/\kappa)^2$.

\begin{figure}[t]
\centering \includegraphics[width=\columnwidth]{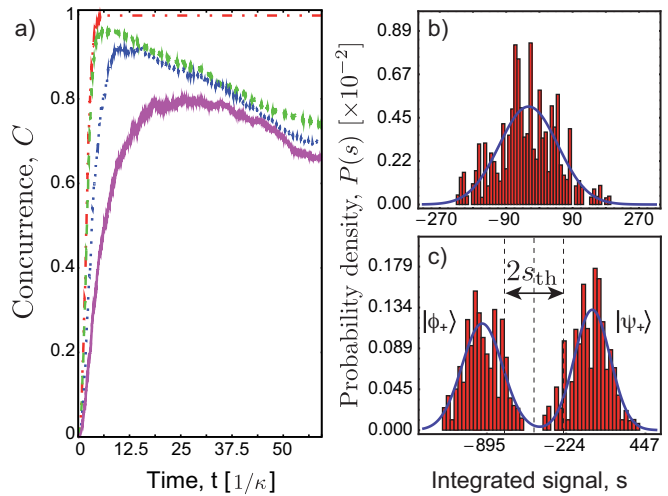}
\caption{(Color online)
a) Concurrence as a function of time for $ 10^4$ trajectories. Dotted-dashed red line: $(\gamma_{1j},\gamma_{\phi j},\eta) = (0,0,1)$, dashed green line: ($\kappa/250$,0,0.8), dotted blue line: ($\kappa/250$,0,0.2) and full purple line: ($\kappa/250$,0,0.05). $\phi=\pi/2$. All other parameters are the same as in Fig.~\ref{fig:fidelities}. b-c) Histograms of the integrated current $s(t)$ at b) $t=1.6/\kappa$ and c) $t=6.3/\kappa$.}
\label{fig:concurrences-histograms}
\end{figure}

To show that the system collapses to $\ket{\psi_+}$ or $\ket{\phi_+}$, and that entanglement is generated with unit probability, Fig.~\ref{fig:concurrences-histograms}a) shows the mean concurrence $E[C(\rho_J)]$, averaged over $ 10^4$ trajectories.  There, departure from unit concurrence in the dotted-dashed red line (no damping, unit detector efficiency $\eta=1$) is only due to measurement-induced dephasing.  The dashed green, dotted blue and full purple lines take into account relaxation with $\kappa/\gamma_{1j}\sim 250$,  and detection efficiency of $\eta=4/5, \eta=1/5, \eta=1/20$, respectively. The latter corresponds to current experimental values~\cite{houck:2008a}. The ratio $\kappa/\gamma_{1j}$  is slightly out of reach of current experiments when taking into account that $\chi^j=10\kappa$ is also required. This cannot be achieved with transmons as current experiments have reached the maximal possible coupling~\cite{devoret:2007a}.  However, new ideas to increase the qubit-cavity coupling can help in achieving these parameters~\cite{bourassa:2009a}.

Small detection efficiency reduces the ratio $\Gamma_{11}/\Gamma_d^{ee,gg}$, which in turns corrupts $\ket{\psi_+}$. As illustrated in Fig.~\ref{fig:concurrences-histograms}a), this results in lower concurrences when $\eta < 1$.  Interestingly, $\ket{\phi_+}$ is not affected by this detection efficiency~\cite{Hutchison:2008}. Nevertheless, improvement in detection efficiency is required to match concurrences that can be realized with an entangling Hamiltonian~\cite{DiCarlo:2009}. Recent improvements with near quantum-limited amplifiers are a good step in this direction~\cite{castellanos-beltran:2008a}.

\begin{figure}[t]
\centering \includegraphics[width=\columnwidth]{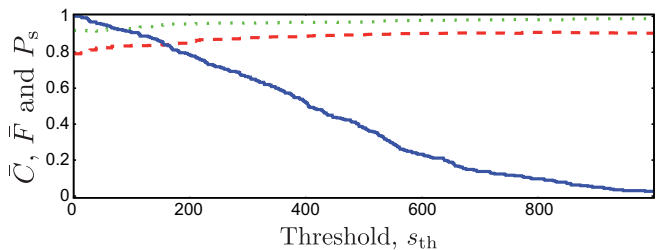}
\caption{(Color online)
Average  concurrence (dashed red line), average fidelity to  $\ket{\phi_+}$ and $\ket{\psi_+}$ (dotted green line)  and success probability (full blue line) as a  function of the threshold at time $t=18.5/\kappa$.  $(\gamma_{1j},\gamma_{\phi j},\eta)$ = ($\kappa/250$,0,0.05). The other parameters are the same as in Fig.~\ref{fig:fidelities}. The ac-Stark shift affecting  $\ket{\psi_+}$ has been corrected before evaluating the fidelity. }
\label{fig:threshold}
\end{figure}

Having generated one of the two orthogonal entangled state, it is necessary to distinguish them efficiently. Using the experimental record $J(t)$ to compute
$\rho_J(t)$ from the SME Eq.~\eq{eq:SME} is not efficient since the record is widely fluctuating.  As a result, a useful and more efficient quantity to distinguish the states is the integrated current
\be
s(t) = \sqrt{\Gamma_{11}^s} \int_0^tJ(t')dt',
\ee
where $\Gamma_{11}^s$ is the steady-state value of $\Gamma_{11}(\pi/2)$. Fig.~\ref{fig:concurrences-histograms}b) and c) show two histograms of $s(t)$ at times $t=1.6/\kappa$ and $t=6.3/\kappa$. These results are for $\eta=1$ and exclude damping for illustration purposes. The full blue lines are Gaussians fits to the histograms.  These separate at a rate $\sim \Gamma_{11}$.  At times large compared to $1/\Gamma_{11}$, but short compared to $T_1$ and $T_2$, the distributions are well separated and correspond to $|\psi_+\rangle$ and $|\phi_+\rangle$.

As shown in Fig.~\ref{fig:concurrences-histograms}c), we introduce a threshold $s_\mathrm{th}$ to distinguish these states.  All outcomes with $s(t)<s_0-s_\mathrm{th}$ (condition $c=-$) are assigned to $\ket{\phi_+}$, while those with $s(t)>s_0+s_\mathrm{th}$ (condition $c=+$) to $\ket{\psi_+}$ where $s_{0}$ is the median of $s$.  Values outside this range are disregarded. A success probability $P_\mathrm{s}$ can then be defined as the probability for $s$ to be outside the range $s_0\pm s_\mathrm{th}$. To quantify the success in generation and distinguishability of the entangled states, we define the average fidelity  $\bar F = [\bra{\phi_+}E_{-}[\rho_J]\ket{\phi_+}+\bra{\psi_+}E_{+}[\rho_J]\ket{\psi_+}]/2$ and average concurrence $\bar C= [C(E_{+}[\rho_J])+C(E_{-}[\rho_J])]/2$. $E_{c}[\rho_J]$ represents the ensemble average over $\rho_J$ for condition $c=\pm$. These quantities are illustrated as a function of $s_\mathrm{th}$ for the fixed integration time $t=18.5/\kappa$ in Fig.~\ref{fig:threshold}. Even when keeping all events ($s_\mathrm{th} = 0$), $\bar F$ and $\bar C$ are large with values $0.92$ and $0.79$ respectively. That is, with this procedure, it is possible to create and distinguish highly entangled states with unit probability. If  willing to sacrifice some events, this average fidelity and concurrence is increased to 0.98 and 0.91,  respectively. The deviation from unity in the large $s_\mathrm{th}$ limit is due to slight corruption of the state $\ket{\psi_+}$ discussed previously.

In conclusion, we have shown how measurements in the dispersive regime of two-qubit cavity QED can be tuned from accessing single to multi-qubit information, thus allowing for example  parity measurements.  In addition to allowing complete characterization of the two-qubit states~\cite{filipp:2008a,DiCarlo:2009,chow:2009a} and the implementation of quantum information protocols~\cite{louis:2007a,bacon:2006a,kerckhoff:2009a}, this allows for generation of entanglement by measurement with unit probability.

\begin{acknowledgments}
We thank D. Poulin and M.~P. da Silva for valuable discussions.  KL was supported by FQRNT and NSERC;  JMG by a CIFAR Junior Fellowship, MITACS, MRI and NSERC; AB by NSERC, CIFAR and the Alfred P. Sloan Foundation.
\end{acknowledgments}


\end{document}